%% file: manuscript.tex
\documentclass{article}
\usepackage{spconf,amsmath,graphicx}

\usepackage{enumitem}
\setlist{nosep, leftmargin=14pt}

\usepackage{mwe} 
\usepackage{tabularray}
\usepackage{url}
\usepackage[hidelinks]{hyperref}
\usepackage{balance}

\title{Evaluating the Impact of Sequence Combinations on Breast Tumor Segmentation in Multiparametric MRI}
\name{Hang Min\sthanks{Corresponding author. Email: hang.min@csiro.au}$^{1}$
\quad Gorane Santamaria Hormaechea$^{2}$ \quad Prabhakar Ramachandran$^{3}$
\quad Jason Dowling$^{1}$}

\address{$^{1}$ Australian e-Health Research Centre, CSIRO, Herston, Australia \\
         $^{2}$ Department of Radiology, Princess Alexandra Hospital, Woolloongabba, Australia \\
         $^{3}$ Department of Radiation Oncology, Princess Alexandra Hospital, Woolloongabba, Australia}

\begin{document}
%
\maketitle
\begin{abstract}
Multiparametric magnetic resonance imaging (mpMRI) is a key tool for assessing breast cancer progression. Although deep learning has been applied to automate tumor segmentation in breast MRI, the effect of sequence combinations in mpMRI remains under-investigated. This study explores the impact of different combinations of T2-weighted (T2w), dynamic contrast-enhanced MRI (DCE-MRI) and diffusion-weighted imaging (DWI) with apparent diffusion coefficient (ADC) map on breast tumor segmentation using nnU-Net. Evaluated on a multicenter mpMRI dataset, the nnU-Net model using DCE sequences achieved a Dice similarity coefficient (DSC) of 0.69 $\pm$ 0.18 for functional tumor volume (FTV) segmentation. For whole tumor mask (WTM) segmentation, adding the predicted FTV to DWI and ADC map improved the DSC from 0.57 $\pm$ 0.24 to 0.60 $\pm$ 0.21. Adding T2w did not yield significant improvement, which still requires further investigation under a more standardized imaging protocol. This study serves as a foundation for future work on predicting breast cancer treatment response using mpMRI.
\end{abstract}
\section{Introduction}
\label{sec:intro}

Breast cancer is one of the most prevalent cancers among women worldwide~\cite{waks2019breast}. Multiparametric magnetic resonance imaging (mpMRI) is a valuable tool in breast imaging that facilitates not only the diagnosis of breast cancer but also the assessment of disease progression and treatment response~\cite{marino2018multiparametric}. Commonly incorporating T2-weighted (T2w), dynamic contrast-enhanced MRI (DCE-MRI) and diffusion-weighted imaging (DWI), mpMRI offers both anatomical and functional information of tumors. Specifically, T2w imaging delivers anatomical and fluid content information, while DCE and DWI quantify tumor perfusion and diffusion patterns respectively~\cite{wahid2022evaluation,hendrick2008breast}. Tumor segmentation is an important step in the quantitative analysis of breast MRI data and is commonly carried out manually by radiologists~\cite{xu2023deep}. However, manual segmentation of mpMRI data is both challenging and labor-intensive, given that mpMRI generates a large volume of data across multiple image sequences and temporal resolutions. 

In recent years, deep learning (DL) techniques have gained significant attention in the field of medical image analysis. They are increasingly applied to automate breast cancer segmentation on breast MRI. Chen \textit{et al.}~\cite{chen2018spatio} introduced a U-Net~\cite{ronneberger2015u} with ConvLSTM structure~\cite{shi2015convolutional} to analyze both spatial and temporal features in DCE-MRI for breast tumor segmentation. However, this approach only operates at the slice level rather than the entire 4D data (which involve 3D volumetric DCE images with a temporal dimension). Both Xu \textit{et al.}~\cite{xu2023deep} and Janse \textit{et al.}~\cite{janse2023deep} utilized nnU-Net~\cite{isensee2021nnu}, a self-adaptive network, to segment breast tumors on DCE-MRI. Despite the prevalence of DL applications in DCE-MRI, there is a noticeable gap in research focusing on tumor segmentation on mpMRI. Zhu \textit{et al.}~\cite{zhu2022development} proposed applying a 2D VNet~\cite{milletari2016v} on DCE-MRI and attention-UNet~\cite{oktay2018attention} on DWI for breast tumor segmentation. However, this work also only operates on 2D slice-level instead of the multimodal 3D image data with the temporal dimension generated by mpMRI. Currently, there remains a lack of studies which investigate how to optimally combine mpMRI sequences to fully harness the rich information they offer for enhancing breast tumor segmentation.

This study aims to investigate the impact of different combinations of anatomical and functional sequences in mpMRI on the performance of breast tumor segmentation. The evaluation was conducted on a publicly available, multicenter mpMRI dataset, using different combinations of T2w, DCE, and DWI sequences with apparent diffusion coefficient (ADC) map as inputs for nnU-Net, U-Net with a ConvLSTM structure, and 3D U-Net. The segmentation outcomes were quantitatively analyzed to identify the most effective sequence combinations.

\section{Methodology}
\label{sec:methods}

\subsection{Datasets}
The Breast Multiparametric MRI for prediction of neoadjuvant chemotherapy (NAC) Response Challenge (BMMR2 Challenge) dataset~\cite{bmmr2_dataset,newitt2019test}, which is a multicenter breast mpMRI dataset, was used in this study. It contains 191 subjects divided into approximately 60\% for training and 40\% for testing. The dataset contains both pre- and post-treatment mpMRI data and only pre-treatment data were used in this study to avoid possible alterations to the lesions induced by NAC. Each subject contains T1-weighted (T1w) DCE sequences ($80s< \text{phase duration} < 100s$ with at least 8 minutes continuous post-injection acquisition), 4 b-value (b=0, 100, 600, 800\(s/mm^2\)) DWI with ADC map and T2w imaging that varies in acquisition sequences and fat suppression methods. One case from the training set and one from the testing set were excluded since not all sequences described in the imaging protocol can be identified. The dataset also includes single-breast ipsilateral cropped DCE images that were used to generate the functional tumor volume (FTV) mask. After the FTV was identified on post-contrast DCE subtraction images, a whole tumor mask (WTM) was subsequently segmented on the ADC map manually. 
\subsection{Data preparation}
To prepare the mpMRI for the DL networks, the T2w, DWI and ADC images were resampled to align with the single breast cropped DCE images. The DCE images were temporally resampled into seven DCE frames. The interval between the pre-contrast and the first post-contrast image was set as 150s, while a 90s interval was maintained among the subsequent post-contrast images. Six DCE subtraction images DCE$_{sub}=\{DCE_{i}-DCE_{0}| i=1, 2, ...6\}$ were then generated for the DL segmentation networks, where $DCE_{0}$ stands for the pre-contrast image and $DCE_{i}$ for the post-contrast ones. 

\subsection{Segmentation model training and evaluation}
The nnU-Netv2, a recent release of the nnU-Net~\cite{isensee2021nnu}, was trained to segment FTV and WTM separately. The nnU-Net is a self-configuring approach for biomedical image segmentation that includes preprocessing, network architecture, training, and post-processing. It has been applied to a wide range of biomedical image segmentation tasks and demonstrated outstanding performance~\cite{alqaoud2022nnUNet}. In this study, the 3D full resolution nnU-Net with multichannel input was used. 

To segment the FTV and WTM, various mpMRI sequence combinations were explored as listed in Table~\ref{table:sequence_comb}. For FTV segmentation, the initial combination started with using all DCE$_{sub}$ images, and then adding T2w. For WTM segmentation, the ADC map, which was used as the primary reference during WTM manual segmentation, was the initial input explored for nnU-Net. Other inputs including DWI, predicted FTV (FTV$_{pred1}$ and FTV$_{pred2}$ as shown in Table~\ref{table:sequence_comb} row 1 and 2) and T2w were subsequently introduced. The nnU-Net models were trained via 5-fold cross validation within the training set for 100 epochs with a batch number of 2 on two NVidia Tesla P100 with 16GB memory.

A 3D U-Net with ConvLSTM structure (ConvLSTM-UNet) and a 3D U-Net with multichannel inputs were used for FTV segmentation on DCE$_{sub}$ images and WTM segmentation on the DWI sequence with ADC map respectively for comparison. The ConvLSTM-UNet, which is an extension of Chen \textit{et al.}~\cite{chen2018spatio}, attaches a ConvLSTM block with 16 output features to the first decoding layer of a 3D U-Net to enable the network to process 3D DCE$_{sub}$ images across 6 temporal frames. The 3D U-Net consists of an encoding and a decoding path, each containing three layers, and a bridge layer in between. The numbers of features for these layers are 16, 32, 64 and 128 respectively. Both networks were trained via 5-fold cross validation within the training set for 100 epochs with a batch number of 2.

\begin{table}[htbp]
\caption{MpMRI sequence combinations. FTV stands for functional tumor volume and WTM for whole tumor mask.}
 \label{table:sequence_comb}
\begin{center}
\begin{tabular}{ccc}
\hline
& Sequence combinations & Prediction output\\
\hline
     1 & DCE$_{sub}$ &  FTV$_{pred1}$  \\

     2 & DCE$_{sub}$, T2w &  FTV$_{pred2}$  \\

     3 & ADC &  WTM$_{pred1}$  \\

     4 & DWI, ADC &  WTM$_{pred2}$  \\

     5 & DWI, ADC, T2w &  WTM$_{pred3}$  \\

     6 & DWI, ADC, FTV$_{pred1}$ &  WTM$_{pred4}$  \\

     7 & DWI, ADC, FTV$_{pred2}$ &  WTM$_{pred5}$  \\

     8 & DWI, ADC, FTV$_{pred1}$, T2w &  WTM$_{pred6}$  \\

     9 & DWI, ADC, FTV$_{pred2}$, T2w &  WTM$_{pred7}$  \\
\hline
\end{tabular}
\label{tab1}
\end{center}
\end{table}

After model training, the best configuration plan was automatically determined by nnU-Net on the validation sets, which was used to generate FTV and WTM predictions on the test set. Several metrics on the voxel level including precision, recall, Dice similarity coefficient (DSC) and 95\% Hausdorff distance (HD$_{95}$) were used to evaluate the FTV and WTM segmentation performance. For ConvLSTM-UNet and U-Net,  the segmentation results were obtained by averaging the prediction outcomes from all models.

\section{Results}
\label{sec:results}

\begin{table*}[ht]
\caption{FTV and WTM segmentation performance using nnU-Net, ConvLSTM-UNet, U-Net with difference mpMRI sequence combinations.}
\label{table:performance_metrics}
\centering 
\begin{tblr}{
  colspec = {cccccccc},
  rows = {font=\ninept},
}
\hline
 \SetCell[r=2]{c} & \SetCell[r=2]{c} Targets & \SetCell[r=2]{c} Networks & \SetCell[r=2]{c} Sequence combinations & \SetCell[c=4]{c} Metrics \\
\hline
& & &          & Precision & Recall & DSC & $HD_{95}^{\mathrm{a}}$ \\ 
\hline
1 & FTV & ConvLSTM-UNet & DCE$_{sub}$ &  0.66 $\pm$ 0.19 & 0.67 $\pm$ 0.21 & 0.62 $\pm$ 0.16 & 15.49 $\pm$ 14.08  \\ 
2 & FTV & nnU-Net & DCE$_{sub}$ & 0.72 $\pm$ 0.21  & \textbf{0.75 $\pm$ 0.21}  &  \textbf{0.69 $\pm$ 0.18} & 14.83 $\pm$ 15.34   \\ 
3 & FTV & nnU-Net & DCE$_{sub}$, T2w  & \textbf{0.72 $\pm$ 0.20} & \textbf{0.75 $\pm$ 0.21} & \textbf{0.69 $\pm$ 0.18} & \textbf{13.67 $\pm$ 13.16}  \\ 
4 & WTM & U-Net & DWI, ADC  &  0.57 $\pm$ 0.26 & 0.61 $\pm$ 0.30 & 0.53 $\pm$ 0.24 & 9.93 $\pm$ 11.90 \\ 
5 & WTM & nnU-Net & ADC  & 0.60 $\pm$ 0.29 & 0.59 $\pm$ 0.30 & 0.54 $\pm$ 0.26 & 10.57 $\pm$ 12.40   \\ 
6 & WTM & nnU-Net & DWI, ADC  & 0.61 $\pm$ 0.26 & 0.63 $\pm$ 0.27  & 0.57 $\pm$ 0.24 & 7.70 $\pm$ 7.22  \\ 
7 & WTM & nnU-Net & DWI, ADC, T2w & 0.61 $\pm$ 0.27 & 0.60 $\pm$ 0.31 & 0.56 $\pm$ 0.25 & 8.44 $\pm$ 8.03 \\ 
8 & WTM & nnU-Net & DWI, ADC, FTV$_{pred1}$  & 0.63 $\pm$ 0.25 & \textbf{0.68 $\pm$ 0.22} & 0.60 $\pm$ 0.21 & \textbf{6.75 $\pm$ 7.33} \\ 
9 & WTM & nnU-Net & DWI, ADC, FTV$_{pred2}$  & 0.64 $\pm$ 0.24 & 0.67 $\pm$ 0.23 & 0.60 $\pm$ 0.21 & 7.06 $\pm$ 7.64 \\ 
10 & WTM & nnU-Net & DWI, ADC, FTV$_{pred1}$, T2w  & 0.64 $\pm$ 0.23 & 0.67 $\pm$ 0.23 & \textbf{0.61 $\pm$ 0.20} & 6.84 $\pm$ 6.41 \\ 
11 & WTM & nnU-Net & DWI, ADC, FTV$_{pred2}$, T2w  & \textbf{0.65 $\pm$ 0.22} & 0.66 $\pm$ 0.23 & \textbf{0.61 $\pm$ 0.20} & 7.42 $\pm$ 7.96 \\ 
\hline
\SetCell[c=8]{c}{$^{\mathrm{a}}$$HD_{95}$ was calculated between the true positive regions (i.e. regions that overlap with the groundtruth) and the groundtruth annotation.}
\end{tblr}
\end{table*}

\begin{figure*}[htbp]
  \centering
  \includegraphics[width=\textwidth]{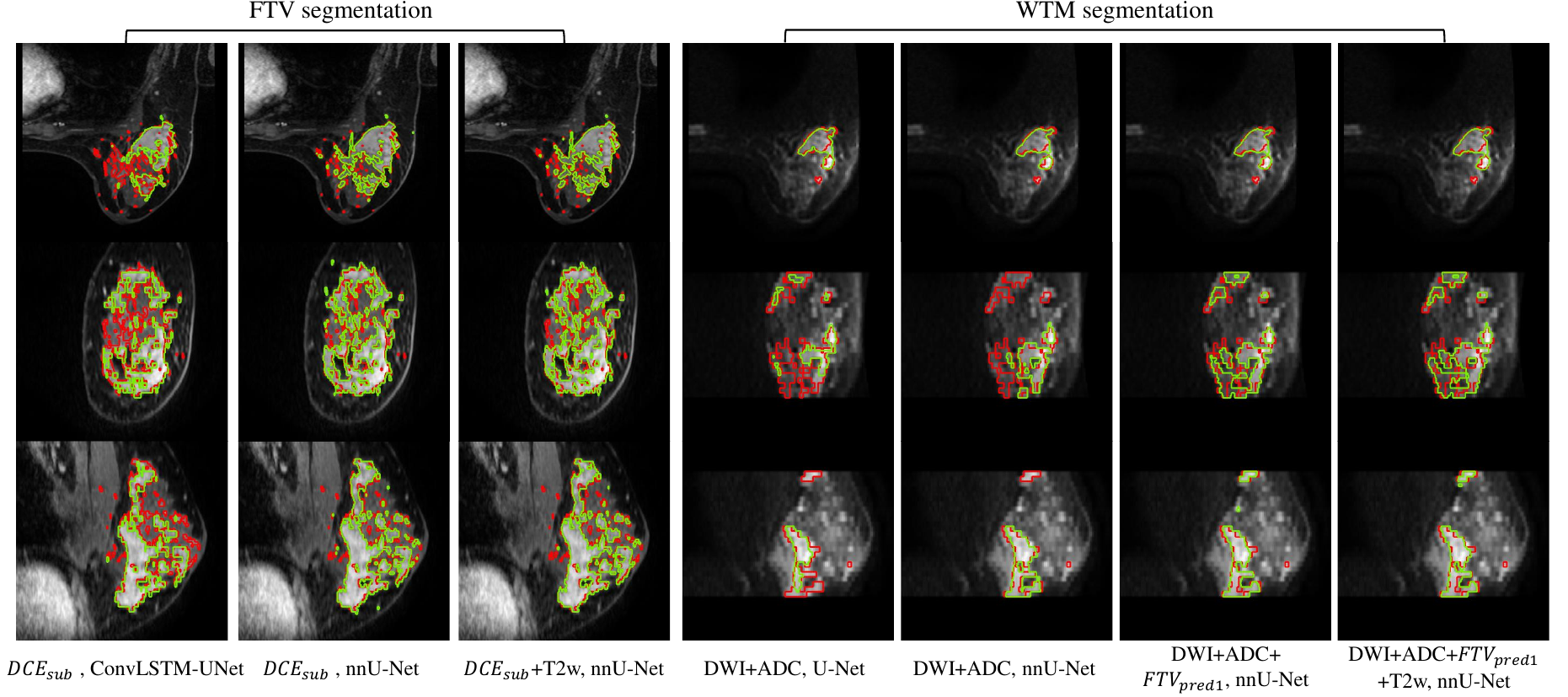}
  \caption{FTV and WTM segmentation examples using ConvLSTM-UNet, U-Net and nnU-Net with different sequence combinations in axial, coronal and sagittal views (from top to bottom). The segmentation examples are outlined on the first post-contrast DCE sequence for FTV segmentation and DWI (b-value=800\(s/mm^2\)) for WTM segmentation. The network segmentation is represented in green and the ground truth in red.}
  \label{fig:seg_examples}
\end{figure*}

Table~\ref{table:performance_metrics} lists the performance of nnU-Net, ConvLSTM-UNet and 3D U-Net using different mpMRI sequence combinations for FTV and WTM segmentation. For FTV segmentation, nnU-Net significantly outperformed ConvLSTM-Unet using DCE$_{sub}$ images (p-value $<$ 0.05 for precision, recall and DSC using Wilcoxon test) as shown in Table~\ref{table:performance_metrics} row 1 and 2. Compared with using only the DCE$_{sub}$ images with nnU-Net, the addition of T2w sequence resulted in similar outcomes (p-value $>$ 0.05 for all metrics) as reflected in Table~\ref{table:performance_metrics} row 2 and 3. 

For WTM segmentation, nnU-Net outperformed U-Net using DWI sequences and ADC map (p-value $<$ 0.05 for precision and DSC) as shown in Table~\ref{table:performance_metrics} row 4 and 6. It can be observed in Table~\ref{table:performance_metrics} row 5, 6 and 7 that incorporating DWI sequences with ADC map yielded better performance than using ADC alone with nnU-Net (p-value $<$ 0.05 for recall and HD$_{95}$), while the addition of T2w sequence did not result in improvement (p-value $>$ 0.05 for all metrics). As indicated in Table~\ref{table:performance_metrics} row 6, 8 and 9, integrating the predicted FTV mask with DWI and ADC map improved the segmentation performance (p-value $<$ 0.05 for DSC and HD$_{95}$), while no significant difference was observed between incorporating FTV$_{pred1}$ and FTV$_{pred2}$ (p-value $>$ 0.05 for all metrics). The addition of T2w, as shown in Table~\ref{table:performance_metrics} row 10 and 11, did not yield a significant difference in performance (p-value $>$ 0.05 for all metrics). The nnU-Net models took approximately 2-3 minutes to segment each image, whereas the ConvLSTM-UNet and 3D U-Net took less than a minute. 

Considering both the segmentation accuracy and model complexity, the optimal approach for segmenting mpMRI appears to be as follows: The nnU-Net models with DCE$_{sub}$ as inputs are firstly utilized for FTV segmentation (generating FTV$_{pred1}$), and the nnU-Net models with DWI+ADC+ FTV$_{pred1}$ as inputs are then employed for WTM segmentation. Fig.~\ref{fig:seg_examples} illustrates the FTV and WTM segmentation examples using ConvLSTM-UNet, U-Net, nnU-Net with several selected sequence combinations.

\section{Discussion}
\label{sec:discussion}

In this study, we assessed the effect of mpMRI input combinations on breast tumor segmentation performance using nnU-Net, which demonstrated superior results compared with ConvLSTM-UNet and 3D U-Net. Our experiments suggest that the most effective practice is to employ DCE$_{sub}$ images for FTV segmentation and a combination of DWI, ADC and the predicted FTV mask for WTM segmentation.

In clinical practice, radiologists commonly employ a combination of T2w, DCE, and DWI along with ADC maps for precise tumor contouring. However, in many studies exploring DL methods for breast tumor segmentation, either the DCE sequence was utilized~\cite{xu2023deep,chen2018spatio,janse2023deep}, or segmentation was conducted independently on DCE and DWI sequences~\cite{zhu2022development}. In this study, the FTV predicted from the DCE sequence was introduced as an additional input channel, along with DWI and ADC map,  to guide the segmentation process. This approach significantly improved the WTM segmentation outcome. The addition of T2w sequence did not result in a significant improvement for FTV and WTM segmentation. This could be attributed to the fact that the imaging protocol for the T2w sequence in the dataset was inconsistent, leading to data variability. Consequently, the potential benefits of incorporating a T2w sequence into the segmentation process cannot be entirely dismissed.

The current study still has several limitations and presents opportunities for further extension. This study only focuses on the pre-treatment mpMRI data and only involved the cropped breast images with tumor provided in the dataset. Segmenting tumors in post-treatment mpMRI poses additional challenges due to potential changes in tumor size, shape, and texture. For future work, we aim to extend our methodology to segment tumors in post-treatment mpMRI, leveraging pre-treatment segmentation as attention guidance to improve the accuracy of post-treatment segmentation. To determine the potential benefits of including T2w sequences in tumor segmentation, further comparative studies should be carried out using an mpMRI cohort with a more standardized imaging protocol. This study also serves as a preliminary investigation into predicting NAC treatment responses. Going forward, we aim to conduct quantitative analysis of breast tumors using mpMRI to identify imaging biomarkers that are predictive of NAC treatment outcomes.

\section{Conclusion}
\label{sec:conclusion}

In this study, we conducted a comprehensive analysis comparing the segmentation outcomes of FTV and WTM on a multicenter breast mpMRI dataset using different sequence combinations. For FTV segmentation, the nnU-Net using DCE subtraction volumes achieved satisfactory results. While the inclusion of the T2w sequence led to slight improvement in some evaluation metrics, these are not statistically significant. For WTM segmentation, the addition of predicted FTV to DWI sequences and ADC map as network inputs significantly improved the segmentation performance. Although the inclusion of T2w did not yield a statistically significant improvement, further experiments are still needed to evaluate its potential for enhancing tumor segmentation performance, particularly under a more consistent imaging protocol. This work serves as a foundational study for future research aimed at predicting breast NAC treatment responses.

\section{COMPLIANCE WITH ETHICAL STANDARDS}

This study used the publicly available BMMR2 dataset~\cite{bmmr2_dataset,newitt2019test} (with CC BY 4.0 license), which is a subset of the ACRIN trial 6698 (NCT01564368). The ethical approval was acquired by the original data providers.

\section{Acknowledgments}
\label{sec:acknowledgments}

This study is supported by Metro South Health Study, Education and Research Trust Account grant (grant number: RSS\_2020\_004), and Australian Government Medical Research Future Fund (grant number: GHFM76722).

\balance

\input{references.bbl}          

\end{document}